\documentclass[twocolumn,showpacs,prb]{revtex4}
\usepackage{graphicx}
\usepackage{epsfig}

\begin{document}

\title{Tunnel effect in ferromagnetic half-metal Co$_2$CrAl-superconductor heterostructures}
\author{E.~M.~Rudenko}
\email{rudenko@imp.kiev.ua}
\thanks{corresponding author}
\author{I.~V.~Korotash}
\author{Y.~V.~Shlapak}
\author{Y.~V.~Kudryavtsev}
\author{A.~A.~Krakovnyi}
\author{M.~V.~Dyakin}

\affiliation{Institute of Metal Physics, National Academy of
Sciences of Ukraine, 252680, Kiev-142, Ukraine}

\begin{abstract}
Ferromagnetic half-metal Co$_2$CrAl films and tunnel contacts
Co$_2$CrAl - insulator (I) - Pb are fabricated and investigated. It
is found that the normalized differential conductivity $\sigma ^{\rm
FS} $ of such tunnel junctions with low resistance is larger than
the normalized differential conductivity $\sigma ^{\rm NS} $ of
known normal metal - I - superconductor type tunnel junctions. It is
shown that the observed increase in $\sigma ^{\rm FS} $ is caused by
the accumulation of spin polarized electrons in a superconductor and
can be used for estimating the spin polarization degree $P$  in
ferromagnets. This method shows that $P$ of L2$_1$-type ordered
Co$_2$CrAl Heusler alloy films at $T = 4.2~{\rm K}$ is close to 1.

~


\end{abstract}


\maketitle

\section{Introduction}

During the recent years spintronics has become a rapidly developing
science. Therefore, studying the peculiarities of spin-polarized
current injection into superconductors (S) is an actual
task.\cite{1,2,3}

Among possible candidates for spin-polarized current injectors from
a ferromagnet (F) into a superconductor some of Heusler alloys (HA)
seem to be more preferable. Indeed, some of the full HA of X$_2$YZ
composition (here X and Y are 3$d$ transition metals, and Z is $s-p$
metal) are ferromagnets and show significant, up to 100 \%, spin
polarization $P$ due to the deep minimum in the energy band gap for
the minority spin electrons at the Fermi level ($E_F$).\cite{4,5,6}
In particular, some of the Co-based HA (and Co$_2$CrAl is among
them) are half-metals with high Curie temperature and high magnetic
moment.\cite{7,8,9} Polycrystalline Co$_2$CrAl alloy ingots have
been fabricated and investigated.\cite{10} However, the measured
spin polarisation $P$ was 62\% that was much less than the
theoretical value. It is also necessary to mention that for the spin
injectors, except the high degree of polarisation, their fabrication
in the form of thin films is very important. Recently there has been
a growing interest in investigating the F-S hybrid structures. They
allow us to receive information on the ferromagnet spin
polarisation, understand the influence of the spin polarized current
on a superconducting state, and establish the physical ground for
development of the spin crioelectronics.

In this work we have fabricated quasimonocrystallic Co$_2$CrAl films
and the tunnel junctions of half-metal ferromagnet Co$_2$CrAl
(HMF)/insulator (I) /Pb (S), {\it i.e.} HMF/I/S structures and
investigated the effect of injecting the spin-polarized electron
current (SPE-current) into superconductor on its superconducting
state.

\section{Experiment}

The tunnel HMF/I/S junctions have a cross-like shape and $200 \times
200 \; \mu {\rm m} ^2 $ junction area. The spin-injectors, L2$_1$ -
type ordered Co$_2$CrAl alloy layers of about 100 nm in thickness,
was deposited first by flash evaporation onto sapphire substrates
kept at different temperatures in a vacuum better than $2 \times 10
^{-5} \; {\rm Pa} $. An insulating barrier layer on the top of the
Co$_2$CrAl films was formed by the natural oxidation of Co$_2$CrAl
layers at the ambient conditions. On the top of the insulating layer
a Pb film of 100 - 200 nm in thickness was thermally deposited. The
microscopic structures of Co$_2$CrAl and Pb films were investigated
by selective-area microdiffraction of transmission electron
microscopy (TEM). It was shown that Co$_2$CrAl and Pb films have
L2$_1$- type of structure and $fcc$-type of lattice, respectively
(see Fig. 1). The magnetic properties of such prepared Co$_2$CrAl
films were investigated in a temperature range of 5 - 350 K using
SQUID-magnetometer. Additionally, in-plane magnetic field
dependences of magnetization $M(H)$ were obtained using the
vibrating sample magnetometer. Temperature dependence of
magnetization for Co$_2$CrAl obtained for cooled and measured at 100
Oe magnetic field (see Fig. 2) reveals the Curie temperature 330 K,
i.e. close to that of the bulk sample. \cite{9}

\begin{figure}[h]
\begin{center}
\mbox{ \psfig{file=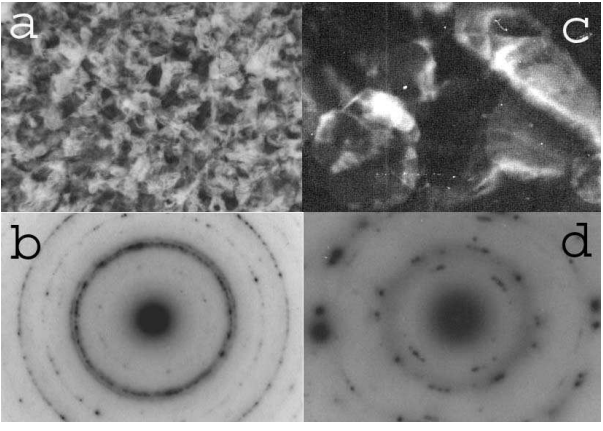,width=8cm}}
\end{center}
\caption{The structure (a,c) and selective-area microdiffraction
(b,d) of TEM for Co$_2$CrAl (a,b) and Pb (c,d) films.}\label{fig1}
\end{figure}

\begin{figure}[h]
\begin{center}
\mbox{ \psfig{file=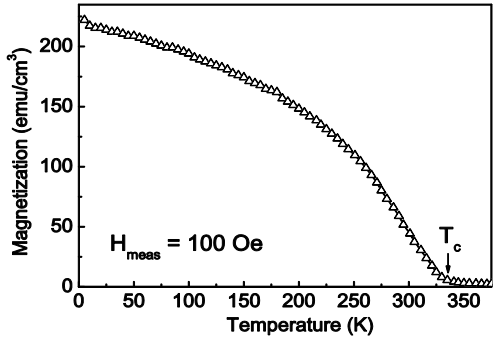,width=8cm}}
\end{center}
\caption{Temperature dependence of magnetization obtained in FC
regime for Co$_2$CrAl films.}\label{fig2}
\end{figure}

Thus fabricated Pb films exhibit the transition into the
superconducting state at $T = 7.2 \; {\rm K} $ with the critical
current density $j _{\rm c} ^{\rm Pb} ( 4.2 \; {\rm K} ) = 3 \times
10 ^6 {\rm A/cm}^2$. Pb as superconductor was chosen for the
following reasons. (1) Pb is the well investigated superconductor
with $s$-type symmetry of the order parameter. (2) It has rather
high critical temperature (7.2 K) and this fact allows us to obtain
high resolution of superconducting parameters at liquid helium
temperatures. (3) Diffusion of Pb into the barrier and the HMF layer
is inhibited because of its rather large ion-radii. (4) Pb is
chemically inactive. \cite{11}

We have found that for these Co$_2$CrAl-I-Pb junctions the value of
normalized conductivity $\sigma ^{\rm FS} \equiv G ^{\rm FS} / G
^{\rm FN} $ essentially differs from the value of fundamental
normalized Giaver conductivity $\sigma ^{\rm NS} \equiv G ^{\rm NS}
/ G ^{\rm NN} $ for tunnel junction of N-I-S type,\cite{12,13,14}
either calculated within the framework of Bardeen-Cooper-Schrieffer
(BCS) theory or determined in experiment. (Here N is normal metal,
$G ^{\rm FS} $ and $G ^{\rm NS} $ - differential conductivities of
the tunnel junction at zero bias, $G ^{\rm FN} $ and $G ^{\rm NN} $
- differential conductivities of the same tunnel junction with S in
the normal state). Besides, we have found that the value of $\sigma
^{\rm FS} $ depends on the value of conductivity $G ^{\rm FN} $,
while the value of $\sigma ^{\rm NS} $ does not depend on $G ^{\rm
NN} $.\cite{11,12,13}

\begin{figure}[h]
\begin{center}
\mbox{ \psfig{file=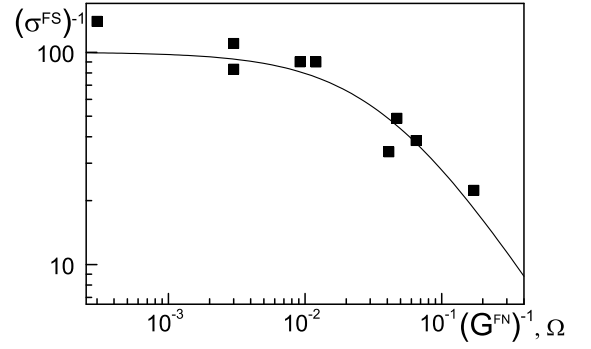,width=8cm}}
\end{center}
\caption{Experimental data (squares) for normalized conductivity
$(\sigma ^{\rm FS})^{-1} $ dependence on conductivity
$(G^{FN})^{-1}$ for a set of Co$_2$CrAl-I-Pb tunnel junctions in the
normal state at $T=4.2~K$, and theoretical dependence (solid line)
for $P = 0.97$, $\Theta \sim (2G^{\rm FN})^{-1}$.}\label{fig3}
\end{figure}

Fig. 3 shows the experimental results for normalized conductivity
$\sigma ^{ \rm FS} $ of Co$_2$CrAl-I-Pb tunnel junctions
investigated at temperature 4.2~K, which shows the aforementioned
effects. One can see that as $(G^{\rm FN})^{-1}$ increases, the
experimental value of $(\sigma^{\rm FS})^{-1}$ changes from 6 up to
100. Calculation within the framework of BCS theory for N-I-Pb
junction at temperature 4.2~K gives $(\sigma^{\rm NS})^{-1} \approx
6.5$. For the N-I-S type of Sn-I-Pb junction it was experimentally
observed that $(\sigma^{\rm NS})^{-1}\approx 5.9$,\cite{15,16} and
for Al-I-Pb junction - $(\sigma^{\rm NS})^{-1}\approx 5.8$.\cite{13}

\section{Discussion}

For N-I-S tunnel junctions dependence of the tunnel current $I$ on
bias $V$ at values $eV$ smaller than the height of the potential
barrier is determined as follows: \cite{14}
\begin{equation}\label{1}
I^{\rm NS} (V)=C\int \limits _{-\infty }^{+\infty }N_{\rm t} (E_{\rm
k} ) [f(E_{\rm k} )-f(E_{\rm k} +eV)]dE_{\rm k},
\end{equation}
where  $C=e^{\rm -1} G_{\rm NN} $,  $G_{\rm NN} =\frac{4\pi e^{2}
}{\hbar } \left|\mathrm{T}\right|^{2} N^{(1)} (0)N^{(2)} (0)$ is
conductance of the junction with both electrodes in the normal
state, $\mathrm{T}$ is the tunneling matrix element, $N^{(1)}(0)$,
$N^{(2)}(0)$ - densities of electronic states at the Fermi level in
the junction banks, $e$ - the elementary charge, $N_{\rm t} (E_{\rm
k} )=Re\frac{\left|E_{\rm k} \right|}{\varepsilon _{\rm k} } $  -
quasiparticles density of states in the superconductor according to
BCS theory, $f(E_{\rm k})$ - Fermi-function of electronic states
distribution with a momentum $k$ and energy $\varepsilon _{\rm k}
=\sqrt{E_{\rm k}^2 -\Delta ^2} $  for a superconductor with the
energy gap $\Delta$.

At $eV \ll  \Delta$ the normalized differential conductivity
$\sigma^{\rm NS}$ of N-I-S tunnel junctions can be calculated as
\cite{14}
\begin{equation}\label{2}
\sigma ^{\rm NS} (T)=\frac{(dI/dV)_{\rm S} }{(dI/dV)_{\rm N} } =\int
\limits _{-\infty }^{+\infty }N_{\rm t} (E) \frac{-\partial f(E)}
{\partial E} dE.
\end{equation}
If we replace the normal metal N in an N-I-S junction with a
ferromagnet F, the dependence (\ref{1}) of the current in an F-I-S
tunnel junction will become as follows:
\begin{equation}\label{3}
I^{\rm FS} (V)=\sum \limits _{\rm \sigma }C_{\rm \sigma } \int
\limits _{-\infty }^{+\infty }N_{\rm t} (E_{\rm k} ) [f(E_{\rm k}
)-f(E_{\rm k} +eV)]dE_{\rm k},
\end{equation}
where  $eC_{\rm \sigma } =\frac{2\pi e^{2} }{\hbar } \left| \rm
T\right|^{2} N_{\rm \sigma}^{(1)} (0) N^{(2)} (0)$  - junction
conductivity for separate spin subzone of ferromagnet for normal
state of both electrodes; spin index $\sigma$ runs over the values
+1($\uparrow$) and -1($\downarrow$); $N_{\uparrow}^{(1)}(0)$,
$N_{\downarrow}^{(1)}(0)$ - densities of the electronic states at
the Fermi level in the ferromagnet for separate spin subzone. Then
$\left(C_{\uparrow } +C_{\downarrow } \right)=e^{-1} G^{\rm FN} $ .

Spin polarization degree $P$ of the ferromagnet is equal
to:\cite{17} $P=\frac{N_{\uparrow } (0)-N_{\downarrow }
(0)}{N_{\uparrow } (0)+N_{\downarrow } (0)} =\frac{C_{\uparrow }
-C_{\downarrow } }{C_{\uparrow } +C_{\downarrow } } $ .

For small bias applied to the junction there is a nonequilibrium
quasiparticles distribution function in superconductor $f(E_{\rm
k})$ which can be described by the equilibrium Fermi-function
$f_0(E_{\rm k})$ with the nonequilibrium additive term $\pm \delta
\mu _{\rm F}$ to chemical potential for two spin subsystems: $\mu
_\uparrow = \mu + \delta \mu _{\rm F}$ è $\mu _\downarrow = \mu -
\delta \mu _{\rm F}$.\cite{18} For $eV \ll \Delta$ the value of this
term is much smaller than the energy gap $\Delta$ and linearly
depends on the bias. \cite{19} Thus the charge and the spin currents
in the junction are as follows:
\begin{equation}\label{4a}
\begin{array}{rl}
\left. I^{\rm FS} \right|_{eV \ll \Delta } = (C _{\uparrow}(eV -
\delta \mu _{\rm F}) + C _{\downarrow}( eV + \delta \mu _{\rm F}) )
\sigma ^{\rm NS}(T) =
& \\
= (1 - P {\rm \kappa}) \sigma ^{\rm NS}(T) eV G^{F\rm N} &
\end{array}
\end{equation}
\begin{equation}\label{4b}
\begin{array}{rl}
\left. I_{\rm s}^{\rm FS} \right|_{eV \ll \Delta } = (C
_{\uparrow}(eV - \delta \mu _{\rm F}) - C _{\downarrow}( eV + \delta
\mu _{\rm F}) ) \sigma ^{\rm NS}(T) =
& \\
= (P - \kappa) \sigma ^{\rm NS}(T) eV G^{F\rm N} &
\end{array}
\end{equation}
where  $\kappa =\left. \frac{\delta \mu _{F} }{eV} \right|_{eV \ll
\Delta } $ .

Let us designate  $\alpha \equiv \left. \frac{I_{\rm s}^{\rm FS}
}{I^{\rm FS} } \right|_{eV \ll \Delta } =\frac{p- \kappa }{1-p
\kappa } $ . Then, taking into account (\ref{2})-(\ref{4b}), the
normalized conductivity of F-I-S junction $\sigma ^{\rm FS}$ can be
calculated as follows:
\begin{equation}\label{5}
\begin{array}{rl}
\sigma ^{\rm FS} (P,T) = \left. \frac{1}{G ^{\rm FN} } \frac{I ^{\rm
FS} }{V} \right|_{eV \ll \Delta } = (1 - p \kappa ) \sigma ^{\rm NS}
(T) = & \\
= \frac{(1-P^{2} )}{1 - \alpha P} \sigma ^{\rm NS} (T) &
\end{array}
\end{equation}

From dependence (\ref{5}) we see that the value of the normalized
tunnel F-I-S junction conductivity $\sigma ^{\rm FS} $ depends on
the ferromagnet spin polarization degree $P$ and the value $\kappa $
in superconductor (or ratio $\alpha$ of the charge current value
$I^{\rm FS}$ and the spin current value $I_s^{\rm FS}$). So the
value of $\sigma^{\rm FS}$ can differ essentially from the N-I-S
tunnel junction normalized conductivity $\sigma ^{\rm NS} $
calculated according to BCS theory or measured in experiments.

It is known that the presence of excess quasiparticles in N-I-N
tunnel junctions results in insignificant (about few percent)
increase in differential resistance at zero bias. \cite{20} The
observed effect is connected with the excess quasiparticles
occupying the initially free energy states, which reduces tunneling
probability for electrons. The quantity of excess quasiparticles is
connected with time of their relaxation on low-energy phonons, whose
density is insignificant.

Aronov\cite{21} has theoretically shown that the tunneling current
in ferromagnet-superconductor junction leads to spin polarization of
quasiparticles in a superconductor. Both the spin polarization of
quasiparticles in the superconductor and the external injection of
spin polarized current leads to accumulation of the excess
nonequilibrium spin polarized quasiparticles. The physical reason of
this phenomenon is connected with the fact that spin polarized
electrons can not directly recombine into singlet Cooper pairs. For
them to recombine, the electron spin flip processes are preliminary
required. The probability of such processes in the superconductor in
the absence of the magnetic impurities is extremely small.\cite{21}
Presence of the excess quasiparticles in the superconductor in
addition to the thermal ones can result in lower conductance of the
tunnel junction.

In our investigated F-I-S Co$_2$CrAl-I-Pb film structures in the
superconducting Pb there are no possibilities for effective
flip-processes of the injected spin-polarized quasiparticles. That
is why there is a possibility of accumulation of nonequilibrium
spin-polarized electrons and their occupying the initially free
energy levels in the superconductor, and, as a result, blocking of
tunneling process from the ferromagnet. This will reduce the
conductivity of the tunnel contact $\sigma^{\rm FS}$.

Reduction of $\sigma^{\rm FS}$ in comparison with $\sigma^{\rm NS}$
due to the spin blocking of the tunneling process will take place
only if the effective life-time for the spin-flip processes
$\tau_{\rm sf}$ for the spin-polarized electrons  in the junction
region is longer than the electron life-time at a given temperature
for the tunneling through the barrier $\tau _{_{T}} \sim
\frac{1}{v_{k_{\bot }} P(v_{k_{\bot }} )}$, where $P(v_{k_{\bot }} )
$ is probability of the electron transfer through a barrier,
$v_{k_{\bot }} $ - component of the electron's velocity normal to
the barrier. \cite{14}

The ratio of the number of electrons which undergo the spin-flip
process to the total number of electrons, which pass through the
barrier into the superconductor equals $\tau_{_{\rm T}} / \tau _{\rm
sf}$. Spin-flip doubles the portion of electrons capable to
recombine. It will increase the number of free energy levels and
will reduce spin polarization degree of excess quasiparticles in a
superconductor. Magnitude of spin depolarization effect due to this
mechanism can be characterized by the {\it factor of recombination
of spin depolarization} $\Theta = 2 \tau_{_{\rm T}} / \tau _{\rm
sf}$. Parameter $\Theta$ defines the portion of electrons which
recombine into singlet Cooper pairs due to the spin-flip as a part
of difference between the number of electrons which have tunneled
into a superconductor with major and minor spin projections.

Multiplier ($P - \kappa $) in (\ref{4b}) gives the decrease of
$\left. I_{\rm s}^{\rm FS} \right| _{eV \ll \Delta } $ with increase
of $\kappa $. On the other hand, this value is equal to $\Theta P$.
Equating the two expressions, one can obtain: $ \kappa = P(1 -
\Theta) $, $ \alpha = \frac {p - \kappa }{1 - p \kappa } = \frac {p
\Theta }{1 - p ^{2} (1 - \Theta )} $ and
\begin{equation}\label{7}
\sigma ^{\rm FS} (p,T) = (1 - p ^{2} (1 - \Theta )) \sigma ^{\rm NS}
(T)
\end{equation}

In Fig. 4 it is shown a set of lines of possible $( \sigma ^{\rm FS}
/  \sigma ^{\rm NS} )^{-1} $ values in F-I-S junction depending on
$P$ at different values of $\Theta$.

In the case with zero spin current ($\alpha = 0$): \cite{16}  $
\sigma ^{\rm FS} =  \sigma ^{\rm NS} (1 - P ^2) $. If in this case
$P = 1$, then $ \sigma ^{\rm FS} =  \sigma ^{\rm NS} (1 - P ^2) =
0$, {\it i.e.} injection current is absolutely blocked.

\begin{figure}[h]
\begin{center}
\mbox{ \psfig{file=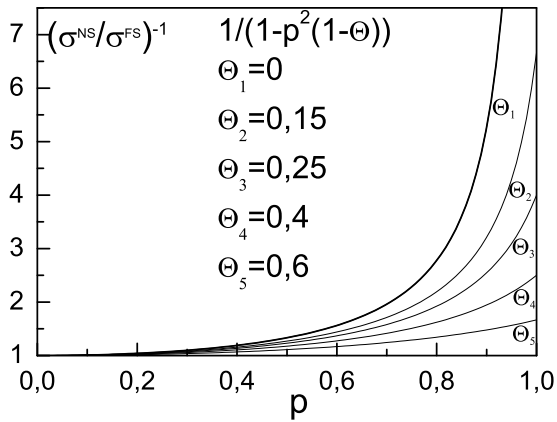,width=8cm}}
\end{center}
\caption{Dependence of the normalized conductivity of F-I-S junction
$\sigma ^{\rm FS}$ on the spin polarization degree $P$ for different
values of the recombination of spin depolarization parameter $\Theta
$ and for different values of $\alpha $.}\label{fig4}
\end{figure}

The obtained theoretical results can be used for experimental
determination of spin polarization degree in ferromagnets.

For two different F-I-S junctions with the same ferromagnet
dependence (\ref{7}) gives  $\sigma _{1,2}^{\rm FS} (p,T) = \sigma
^{\rm NS} (T)(1-p ^{2} (1 - \Theta _{1,2} ) ) $  (we assume that
$\sigma _{1}^{\rm FS} < \sigma _{2}^{\rm FS} $ and, correspondingly,
$\Theta _1 < \Theta _2$). Having determined the values of the
normalized conductivities $\sigma _{1,2}^{\rm FS} $ of these
junctions from the experiment, one can determine spin polarization
degree of this ferromagnet as follows:
\begin{equation}\label{8}
1-P^2=\frac{\sigma _{1}^{\rm FS} }{\sigma ^{\rm NS} (T)}
\frac{\frac{\Theta _{2} }{\Theta _{1} } -\frac{\sigma _{2}^{\rm FS}
}{\sigma _{1}^{\rm FS} } }{\frac{\Theta _{2} }{\Theta _{1} } -1} =
\frac{\sigma _{1}^{\rm FS} }{\sigma ^{\rm NS} (T)}
\frac{\frac{G_{1}^{\rm FN} }{G_{2}^{\rm FN} } -\frac{\sigma
_{2}^{\rm FS} }{\sigma _{1}^{\rm FS} } }{\frac{G_{1}^{\rm FN}
}{G_{2}^{\rm FN} } -1},
\end{equation}
where accounted $\frac{\Theta _{2} }{\Theta _{1} } = \frac{\tau
_{_{\rm T2}} }{\tau _{_{\rm T1}} } = \frac{P _{1} (v _{\rm k _{\bot
}} )}{P _{2} (v _{\rm k_{\bot }} )} = \frac{G _{1}^{\rm FN}
}{G_{2}^{\rm FN} } $.\cite{14}


We have measured the values of $G^{\rm FN}$ and $\sigma^{\rm FS}$
for different Co$_2$CrAl-I-Pb junctions formed with the films of the
same half-metal ferromagnet Co$_2$CrAl (Fig. 1). Using different
pairs of values  $G_{\rm i}^{\rm FN}$ and $\sigma_{\rm i}^{\rm FS}$,
we have determined the spin polarization degree $P$ of this
half-metal ferromagnet Co$_2$CrAl and obtained $P = 0.97 \pm 0.03$.
In (\ref{8}), we used calculated from the BCS theory value $(\sigma
^{\rm NS}) ^{-1} = 6.5 $ for temperature $T = 4.2\; {\rm K} $.
Theoretical curve (\ref{7}) for determined $P = 0.97$ and $\Theta
\sim (2 G ^{\rm FN}) ^{-1} $ is shown in Fig. 3 and is in good
agreement with the experimental data.

As we see, the value of spin polarization degree of the
ferromagnetic half-metal Co$_2$CrAl is $P=0.97 \pm 0.03$, that only
slightly differs from the theoretical value $P_{\rm t}=1$.

\section{Conclusions}

1.  The phenomenon of spin blocking of the tunnel current in
Co$_2$CrAl-I-Pb junctions, which leads to a change in the normalized
differential conductivity of the junctions at zero bias is observed.
The value of spin blocking depends on the ferromagnet spin
polarization degree $P$ and the value of the factor of recombination
of spin depolarization $\Theta = 2\tau_{\rm T} / \tau_{\rm sf}$.

2.  It is established that the normalized conductivity $\sigma ^{\rm
FS}$ of Co$_2$CrAl-I-Pb tunnel junctions can be essentially smaller
than the normalized conductivity $\sigma ^{\rm NS}=0.15$ of N-I-S
tunnel junction. The value of $\sigma ^{\rm FS}$ depends on the
ferromagnet spin polarization degree $P$ and the value of junction
conductivity in the normal state $G^{\rm FN}$. It is revealed that
$\sigma ^{\rm FS}$ can be smaller than 0.01 for tunnel junctions
made of quasimonocrystal films of ferromagnetic half-metal Heusler
alloy Co$_2$CrAl.

3.  We have fabricated quasimonocrystal films of ferromagnetic
half-metal Heusler alloy Co$_2$CrAl with the spin polarization
degree $P=0.97 \pm 0.03$, that is close to the theoretical value
$P_{\rm t}=1$.

4.  It is shown that measuring the differential conductivity in
tunnel junctions of ferromagnet-insulator-superconductor type at
zero bias allows us to determe the ferromagnet spin polarization
degree.

\end{document}